\documentstyle[12pt,epsf]{article}

\begin{document}

\noindent
{\large\bf Magnon Broadening Effects in Double Layered Manganite
La$_{1.2}$Sr$_{1.8}$Mn$_2$O$_7$}

\vspace{0.5cm}
\noindent
Nobuo Furukawa$^a$, Kazuma Hirota$^b$

\vspace{0.5cm}
\noindent
$^a$ Department of Physics, Aoyama Gakuin University,\\
Setagaya, Tokyo 157-8572, Japan\\
$^b$ CREST,Department of Physics, Tohoku University,
Sendai 980-8578,  Japan

\vspace{1.5cm}
\noindent
{\bf Abstract:}\\
Magnon linewidth of La$_{1.2}$Sr$_{1.8}$Mn$_2$O$_7$ 
near the Brillouin zone boundary is investigated from
both theoretical and experimental points of view.
Abrupt magnon broadening is ascribed to a strong magnon-phonon coupling. 
Magnon broadening observed in cubic perovskite manganites
is also discussed.


\section{Introduction}

Magnetic excitation spectra of colossal magnetoresistance (CMR) manganites
in the ferromagnetic metal phase attract our attention in the point whether
they can be understood by the conventional double-exchange (DE) mechanism.
For (La,Sr)MnO$_3$ and (La,Pb)MnO$_3$ where $T_{\rm c}$ is relatively high,
a cosine-band type magnon dispersion 
is observed \cite{Perring96,Martin96,Moudden98}. At low temperature,
Magnon linewidth $\Gamma$ is
narrow enough throughout the Brillouin zone, which makes it
possible to observe well-defined magnon branches,
and it becomes broad at finite temperature.
The DE model explains the cosine-band dispersion \cite{Furukawa96}
as well as the temperature dependence of the linewidth 
in the form $\Gamma \propto (1-M^2)\, \omega_q$,
where $M$ is the magnetization normalized by the saturation value
 and $\omega_q$ is the magnon
dispersion \cite{Furukawa98}.
The origin of the magnon broadening is the Stoner absorption,
which disappears at $T\to0$ (or $M\to 1$) due to the half-metallic nature
of the system.

For compounds with lower $T_{\rm c}$,
 Doloc {\em et al.}\ \cite{VasiliuDoloc98} observed 
broadening of magnon dispersion.
They claimed that the abrupt increase of linewidth near the zone boundary 
can 
not be explained by DE mechanism alone.
One of the possible explanations is that 
the broadening  is caused by
the magnon-phonon interaction \cite{Furukawa99bx}.
A strong coupling between magnons and phonons
are through the modulation of the exchange coupling
by the lattice displacement.

Anomalous broadening of magnon linewidth is also observed in the
double-layered manganite  La$_{1.2}$Sr$_{1.8}$Mn$_2$O$_7$ \cite{Fujioka99}.
Intra double-layer coupling creates optical and acoustic branches
of magnons. Two-dimensional dispersion of both branches 
indicates that the inter double-layer coupling is sufficiently weak.
Magnon broadening near the zone boundary is also observed in this compound.
In this paper we investigate the possibility of
 this broadening caused by the magnon-phonon interaction.

\section{Comparison between theory and experiment}

As for dispersionless optical phonon 
with frequency $\Omega_0$, the magnon linewidth 
due to magnon-phonon interaction is
given by $\Gamma(q) \propto D(\omega_q - \Omega_0)$,
where $D(\omega)$ is the magnon density of states \cite{Furukawa99bx}.
In a two dimensional system, we have step-function like behavior
\begin{equation}
 \Gamma(q) = \left\{ 
                \begin{array}{ll}
                        \Gamma_0 \qquad & \omega_q > \Omega_0 \\
                        0        & \omega_q < \Omega_0
                \end{array}
             \right.   .
\end{equation}
When  a magnon  with momentum $q$ has energy $\omega_q >\Omega_0$,
it is possible to find an elastic channel to decay into
a magnon-phonon pair with momentum
$q'$ and $q-q'$, respectively,
which satisfies $\omega_q = \omega_{q'} + \Omega_0 $.
This is the reason why magnon linewidth abruptly becomes 
broad as magnon branch crosses that of the phonon.

Let us now compare the theoretical results with experimental data.
We show  inelastic neutron scattering
intensities for La$_{1.2}$Sr$_{1.8}$Mn$_2$O$_7$ 
in Fig.~1, where a contour map is plotted
in the $\omega$-$q$ plane. Scattering vector is taken as
$(1+q,0,5)$ in the reciprocal lattice units. Details of experimental are 
given
in ref.~\cite{Fujioka99}.
A well-defined acoustic magnon branch is observed near the zone center.
We also see optical phonon which is nearly dispersionless at 
$\omega\sim 20{\rm meV}$.  
Above $q \sim 0.3$ where magnon branch and phonon branch crosses,
we see an abrupt increase of the magnon linewidth.
A weak trace of the dispersion is observed above the crossing point.

The data is consistently explained as follows.
Magnon dispersion is cosine-band like with the zone boundary
energy  $\sim 40 {\rm meV}$,
which crosses with the optical phonon with $\Omega_0\sim 20{\rm meV}$.
A strong coupling between magnons and phonons creates
abrupt magnon broadening above the crossing point.

\section{Discussion}

Magnon dispersions so far observed in the ferromagnetic metal phase
of manganites are 
well defined near the zone center regardless of compounds and 
dimensionalities.
Zone boundary broadening is, however, strongly compound dependent.
The present result suggests that the zone-boundary magnon broadening
is influenced by the strength of the magnon-phonon interactions.
Although magnon-phonon dispersion 
crossing is also reported in three-dimensional
manganites \cite{Moudden98,Dai99x}, zone-boundary broadening is
observed only in low $T_{\rm c}$ compounds. This implies a
relation between $T_{\rm c}$ and spin-lattice interaction strength.
Strong damping of the zone-boundary magnons might also
explain the ``zone-boundary softening'' of magnons in low $T_{\rm c}$
manganites \cite{Hwang98}, if we assume that the zone-boundary
flat dispersion observed by neutron inelastic scattering
is allocated as an optical phonon branch, while the real zone-boundary
magnon branch at higher frequency
is wiped out above the magnon-phonon crossing point.

Further detailed studies of
the relations between the magnon linewidth broadening
above the magnon-phonon crossing point and the other magneto-elastic
behaviors  will clarify the role of the spin-lattice interactions
to various physical properties.

N.F. thanks J. Fernandez-Baca for discussion. K.H. acknowledges H. Fujioka, 
M. Kubota, H. Yoshizawa, Y. Moritomo and Y. Endoh for experimental
collaborations.
This work is partially supported by Mombusho Grant-in-Aid
for Priority Area.



\noindent
\section*{Figure captions.}

{\bf Figure 1:}\\
 The dispersion relation of acoustic branch of spin wave of
La$_{1.2}$Sr$_{1.8}$Mn$_{2}$O$_{7}$ at 10~K ($I4/mmm$: $a=3.87$~\AA, 
$c=20.1$~\AA).  Measurements were carried out on the
triple-axis spectrometer TOPAN located in the JRR-3M reactor of JAERI. 
PG (002) reflection of pyrolytic graphite was use to monochromate
 and analyze neutrons.  Data
were taken at every 1~meV and 0.05~rlu (reciprocal lattic unit) 
along  $(1+q\ 0\ 5)$ and accumulated
for 7~min.  Contours are drawn every 20 counts between 0 and 400. 
Nearly dispersionless optical phonon branch is also observed at
 $\omega \sim 20$~meV.

\end{document}